\documentclass[mathleft
]{an}
\usepackage{graphicx}
\usepackage{times}
\begin{document}

\def\deg{\ensuremath{^{\circ}}}

\Pagespan{789}{}
\Yearpublication{2006}%
\Yearsubmission{2005}%
\Month{11}%
\Volume{999}%
\Issue{88}%

\title{High Frequency GPS sources in the AT20G Survey}

\author{P.\, J.\, Hancock\inst{1}\fnmsep\thanks{Corresponding author:
  \email{p.hancock@physics.usyd.edu.au}\newline}
}
\titlerunning{HFP sources in the AT20G}
\authorrunning{Hancock}
\institute{
Sydney Institute for Astronomy (SIFA), School of Physics, University of Sydney 2006 NSW Australia
}
\received{30 May 2005}
\accepted{11 Nov 2005}
\publonline{later}


\abstract{The Australia Telescope 20GHz (AT20G) survey was used to select a complete sample of 656 Gigahertz Peaked Spectrum (GPS) sources with spectral turnovers above 5GHz. The AT20G has near simultaneous observations at 4.8, 8.6 and 20GHz, which makes it possible to exclude flat spectrum variability as a cause of a source's peaked spectrum. Optical identification of the sample results in 361 QSOs and 104 galaxies and 191 blank fields. Redshifts are known for 104 of the GPS sources. The GPS sources from the AT20G are discussed and compared to previously known samples. The new sample of high frequency peaking GPS sources is found at a lower redshift than previous samples and to also have a lower 5GHz radio power. Evidence is found to support the idea that the origin of the GPS spectral shape are intrinsically different for galaxies and QSOs. This paper is an elaboration and extension of the talk given at the $4^{th}$ CSS/GPS conference in Riccione in May this year.}

\maketitle

\section{Introduction}
The Australia Telescope 20GHz survey (AT20G) is the first large area sensitive survey at frequencies above 5GHz. A-mong the new and interesting sources detected within the AT20G are the High Frequency Peaking (HFP) subset of Gigahertz Peaked Spectrum (GPS) sources. HFP sources are typified as having a spectral turnover above a few GHz (Dallacasa et al. 2002) and thus far have eluded detailed study, due to the very nature of their spectrum and the lack of sensitive large area surveys above 5GHz. In this paper we discuss a group of HFP sources that were selected from the AT20G survey which are good candidates for such a detailed study. All the GPS sources detected in the AT20G survey have turnover frequencies above 5GHz and are thus HFPs.

\section{AT20G}
The AT20G survey used a wide band analog correlator with three telescopes from the Australia Telescope Compact Array (ATCA) to survey the southern sky at 20GHz. The survey was begun in 2004 and observations were completed in 2007. The survey covers the entire sky south of the equator. The survey was conducted in two stages, a fast scanning stage that imaged the entire sky, and a followup stage that observed candidate sources in order to refine flux and position measurements. The scanning observations were done using the wide band correlator setup, and the data was processed using custom software. The images of the sky that were created from the scanning observations were used to create a list of candidate sources that were then followed up in the second stage of observing. 
Sources that were confirmed at 20GHz were re-observed at 4.8 and 8.6GHz to provide a near simultaneous three point spectrum for each of the sources. Not all sources were able to be observed at the lower frequencies due again to time constraints and also the poor (u,v) coverage available from the ATCA for 4.8 and 8.6GHz at declinations above -15\deg. 

Further details of the AT20G survey and its pilot can be found in Ricci et al. (2004), Massardi et al. (2008), and Hancock et al. (2008 in prep).

\section{Sample Selection}
\begin{figure}[hbt]
\centering
\includegraphics[height=\linewidth, angle=-90, bb= 50 50 555 555]{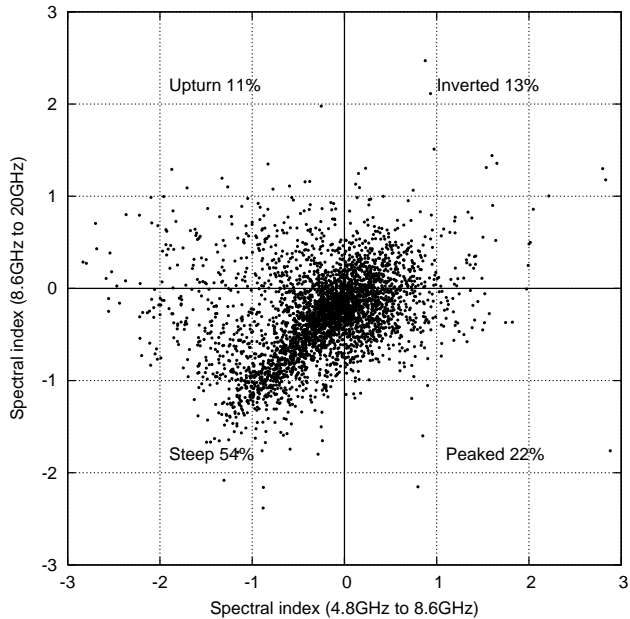}
\caption{Near simultaneous two colour spectral index plot for the 4404 sources that have three frequency measurements in the AT20G. The four regions delineated by the x-y axis are identified by spectral behavior.}
\label{fig:colour-colour}
\end{figure}

The AT20G catalog contains a total of 5450 radio sources. Crossmatching with optical data gives a 90\% identification rate which breaks down into 65\% QSOs, 25\% Galaxies, and 10\% faint or blank fields. Almost all the extragalactic sources are identified with AGN and are unresolved at 20GHz (largest angular size $<$ 5 asec). Figure \ref{fig:colour-colour} shows a two colour plot of the 4404 sources in the AT20G which have three frequency observations. Sources that are inverted between the lower frequencies ($\alpha_{4.8}^{8.6} > +0.2$ where $S=\nu^\alpha$) are considered to be HFP candidates and make up around 20\% of the sample. Sources that are inverted between 8.6 and 20GHz have been found to turn over before 95GHz almost 100\% of the time (Sadler et al. 2008), and are thus included in the GPS sample as sources that have spectral peaks above 20GHz. Sources that are within 2.5\deg  of the Ga-lactic plane are excluded. The resulting list contains 656 GPS sources with peaks above 5GHz.

\section{Additional Information}
\label{sec:additional_information}
Additional information about each GPS source was also collected from SUMSS (280 fluxes at 843MHz, Mauch et al. 2003), NVSS (365 fluxes at 1.4GHz, Condon et al. 1998), NED and 6dF (104 redshifts, Jones et al. 2004), SuperCOSMOS (466 optical identifications and B magnitudes, Hambly et al. 2001).

Optical identifications were taken from the SuperCOSMOS Sky Survey (SSS) class ID. Sources that do not have a SuperCOSMOS class ID were removed from the GPS sample. Redshift information was taken from the NASA/IPAC Extragalactic Database (NED) and are almost exclusively from the 6dF redshift survey.

\section{Sample Properties}
The 104 sources for which there is existing redshift information, were separated into galaxies and QSOs using the class ID flag from the SuperCOSMOS database. As noted in section \ref{sec:comparison} the origin of the GPS spectrum in galaxies and QSOs is intrinsically different and thus the galaxies and QSOs were treated separately. Due to the small number of data points available for the radio spectra of each object the spectral turnover was estimated and binned into one of three categories: 8-15GHz for sources that have a clear peak around the 8.6GHz data point, 15-30GHz for sources that are inverted but show a flattening at 20GHz, and 30GHz+ for sources that are inverted at 20GHz but show no sign of flattening.

\begin{table}
\centering
 \begin{tabular}{lccc}
\hline
          & Galaxies & QSOs  \\
\hline
Number    & 104      & 362  \\
Known z   & 25       &  75  \\
Average z & 0.2      & 1.2   \\
$\log_5(\mathrm{P W/Hz})$ & 24.4 & 26.6  \\
$<\mathrm{B}_{\mathrm{mag}}>$  & $19.4\pm 0.2$ & $19.6\pm 0.1$ \\
$<\mathrm{S}_{20}>$ (Jy) & $0.20\pm 0.04$  &  $ 0.38\pm 0.04$  \\
\hline
 \end{tabular}
 \caption{The properties of the GPS galaxies and QSOs selected from the AT20G survey. The listed errors are errors in the mean.} 
 \label{tab:properties}
\end{table}

\begin{figure}
\vspace{-5mm}
\includegraphics[bb=55 55 555 580, height=\linewidth, angle=-90]{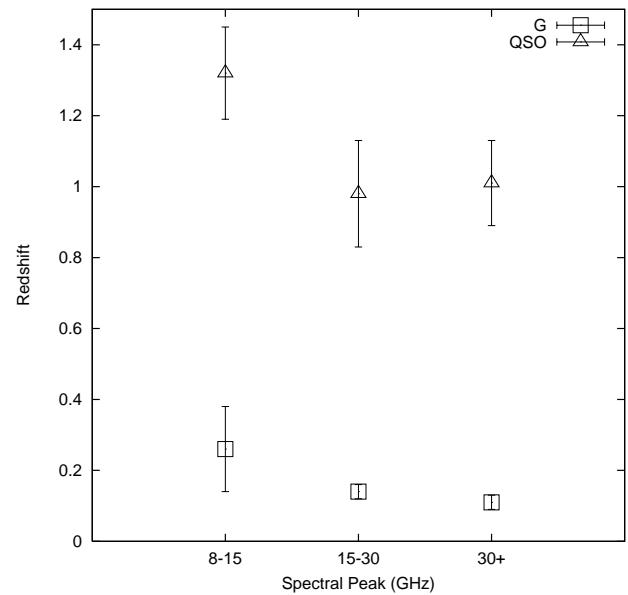}
\caption{Redshift as a function of spectral turnover. Galaxies are shown by squares and QSOs by triangles. }
\label{fig:z_peak}
\end{figure}

\begin{figure}
\includegraphics[height=\linewidth, angle=-90]{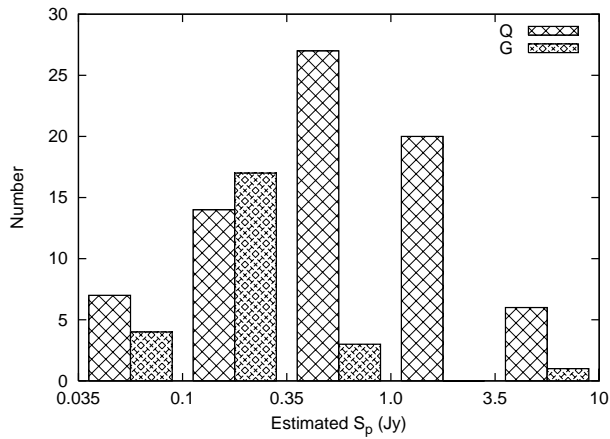}
\caption{Distribution of the flux density at the spectral peak, $\mathrm{S}_\mathrm{p}$, for galaxies and QSOs.}
\label{fig:sp_hist}
\end{figure}

Table \ref{tab:properties} shows the various properties of galaxies and QSOs within the sample. The percentage of galaxies and QSOs that have known redshifts are similar at 20\% and 24\% respectively. As expected the QSOs are found at a systematically higher redshift than the galaxies and have a 5GHz radio power that is 2 orders of magnitude stronger. There is a clear difference between the 20GHz flux density of the galaxies and QSOs with QSOs being found at much higher flux levels on average. If this trend is borne out in future surveys then we can expect to find many more galaxies as the fainter radio sources are probed at 20GHz.

Figure \ref{fig:z_peak} shows the redshift distribution of GPS sources as a function of spectral behavior. In any sample of sources that is defined by a selection frequency, the GPS sources that are found within the sample will have the observed peak closer to the selection frequency. This is because sources that peak further from the selection frequency will appear fainter and are thus less likely to be detected. The lower redshift of the higher frequency turnover sources that is common to both the QSOs and galaxies in figure \ref{fig:z_peak} is consistent with that which is expected from the shape of their spectrum. The separation in redshift between galaxies and QSOs is due to their differing volume density and luminosity. We do not expect to see as many GPS galaxies at high redshift because they are less luminous and thus not as likely to be seen in a flux limited sample of sources. We do not expect to see as many QSOs at low redshift because of their relatively low volume density.

Figure \ref{fig:sp_hist} shows the estimated peak flux density distribution for galaxies and QSOs. As with the spectral peak, not enough spectral information was available to reliably fit the spectrum so the peak flux density was estimated by eye. The distribution of the GPS galaxies and QSOs are quite different, with the galaxies being more sharply peaked around 0.1 to 0.35 Jy estimated peak flux and QSOs being more broadly peaked around 0.35 to 1 Jy estimated peak flux. Each of the GPS sources considered in this paper were classified as either galaxy or QSO by adopting the SSS class ID as stated in section \ref{sec:additional_information}. This classification was done largely automatically and was based on whether or not the sources were resolved on various optical plates and is not influenced by optical magnitude, redshift or any spectral behavior. This clear separation of the two groups of sources in figure \ref{fig:sp_hist} shows that there are two populations of sources, and not just the same population viewed at different distances. 

It should be noted that figure \ref{fig:sp_hist} shows only the sources for which redshifts are known. Whilst the total fraction of sources with known redshifts is the same for galaxies and QSOs, the distribution of known redshifts is not the same within each of the samples. The percentage of QSOs with known redshift increases with $\mathrm{S}_\mathrm{p}$ and may be biasing the peak of figure \ref{fig:sp_hist} in the same direction. The fraction of galaxies with known redshift does not show any such correlation. The width of the two distributions would not be effected by such a bias, only the location of the peak.

\section{Comparison to known GPS samples}
\label{sec:comparison}
Previous GPS studies have contained mainly sources with spectral turnovers below 5GHz. The updated GPS master list of Labiano et al. (2007)\footnote{Online at http://damir.iem.csic.es/$\sim$labiano/GPSlist/GPS\_List.txt, accessed 2008 August 11}(hereafter master list) contains the most complete listing of GPS and HFP sources currently known. The master list consists of 172 sources with 79 of these identified as galaxies and 57 as QSO. A summary of the master list and the sources in this paper is shown in table \ref{tab:compare}.

\begin{table}
\centering
\begin{tabular}{lcc}
\hline
 & Master list & This paper \\
\hline
Typical Peak & 1-5GHz & 8-15GHz \\
Number of sources$^1$ & 172 (136) & 656 (466)\\
Galaxy / QSO percent & 58 / 42 & 23 / 77 \\
Average z & 0.5 / 1.9 & 0.2 / 1.2 \\
Typical $\log\mathrm{P}_5$(W/Hz) &  26.3 / 27.7$^2$ & 24.4 / 26.6 \\
\hline
\end{tabular}
\caption{A comparison of the master list of GPS sources presented by Labiano et al. 2007, and the sources presented in this paper. \footnotesize $^1$Only sources with optical identifications (numbered in parenthesis) are used in the calculations of this table. $^2$Only sources with a listed redshift and 5GHz flux density were included in this calculation (89 in total).}
\label{tab:compare}
\end{table}

The master list contains mostly sources from the northern hemisphere due to the increased availability of large area surveys in the northern hemisphere, however there are 57 southern hemisphere sources present in the master list. Only three of the master lists' southern hemisphere sources have a listed peak above 5GHz (0745-004 at  5.8GHz, 1522-2730 at 5.8GHz, and 1837-7108 at 8.2 GHz). The AT20G survey does not have 4.8 or 8.6GHz observations for sources above -15\deg declination so 0745-004 is not included in the AT20G sample. 1522-2730 is present in both samples and peaks in the 8-15GHz bin. 1837-7108 is seen in the AT20G survey but has a steep spectrum with $\alpha=-0.27$, and is identified with a QSO at redshift 1.35.

While GPS sources are generally the least variable class of radio sources (O'Dea 1998), GPS QSOs are known to have a large amount of variability which can often cause false GPS identification when the spectral observations are not simultaneous (Torniainen et al. 2005). The change in the spectrum of 1837-7108 may be due to the non simultaneous observations used in the initial classification.

Taking the master list as a representative sample of previously known GPS sources we can see that the sources detected in the AT20G survey differ in a number of ways. The typical peak frequency of the AT20G selected sources is much higher than the master list, which is expected since the AT20G will not detect any sources with peaks below 5GHz. The number of sources in the AT20G selected sample is about four times larger than the master list, but the amount of information known about each object is much less than the more well studied sources of the master list. The galaxy to QSO ratio of the two samples are also quite different with the master list having slightly more galaxies than QSOs and the AT20G sample having about three times as many QSOs as galaxies. This change in galaxy to quasar ratio has been noted previously by Stanghellini (2003), among others. The fraction of galaxies in a sample of GPS sources is seen to increase as the turnover frequency decreases even into the Compact Steep Spectrum radio sources that can be thought of as GPS sources with a peak frequency below our detection limit. This supports the idea that the GPS emission from galaxies and quasars are intrinsically different, with the quasar emission originating from the core and that of galaxies from micro lobes and hot spots. The AT20G selected sample of sources are at lower redshift and 5GHz radio power than those in the master list. Galaxies at low radio power and redshift are under represented in current GPS samples and correspond to stages of radio source evolution that precede the currently studied sample of sources. 

\section{Observations at 40 and 95GHz}
In November 2007 a sample of sources from the AT20G catalog were observed at 40 and 95GHz at the ATCA. The observations were a continuation of the 95GHz study of Sadler et al. (2008), and contained a subset of GPS sources from this paper. The GPS sources were selected to be inverted or to be turning over around 20GHz, and to have a 20GHz flux or spectral index that put their predicted higher frequency fluxes within detection limits. In total 21 sources were observed. With an increased knowledge of the source spectra, the estimated spectral turnover frequency and flux density were able to be fitted using the GPS spectrum of Snellen et al. (1998):
\begin{equation}
\mathrm{S}_\nu = \frac{\mathrm{S_p}}{1-e^{-1}}\left(\frac{ \nu}{\nu_\mathrm{p} }\right)^k (1 - e^{-\left(\frac{\nu}{\nu_p}\right)^{l-k}})
\label{eq:one}
\end{equation}
Where: $\nu_\mathrm{p}$ and $\mathrm{S_p}$ are the peak frequency and flux density, and $k$ and $l$ are the optically thick and thin spectral indicies.
\begin{figure}[hbt]
 \vspace{-8mm}
 \includegraphics[height=\linewidth, angle=-90]{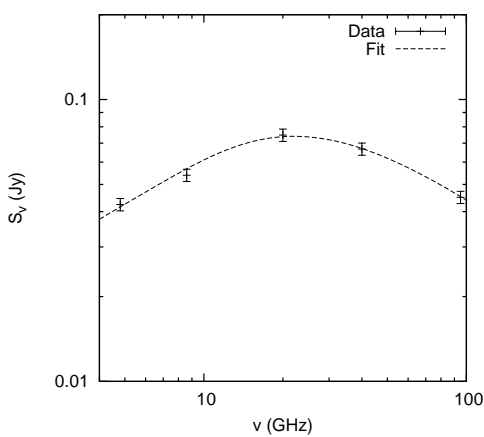}
\caption{The galaxy 1818-5508 at a redshift of 0.07. The dotted line is a fit to equation \ref{eq:one}.}
\label{fig:1818-5508}
\end{figure}

Observations of a number of these sources at the ESO NTT (images and spectra) and Siding Spring Observatory 2.3m telescope (spectra only), as well as data from the literature, are collected and used to probe the dynamics of the radio source. Figure \ref{fig:1818-5508} shows one such source. The results of these observations will be discussed in more detail in Hancock et al. (2009 in prep).

\section{Conclusions and Future Work}
The AT20G survey has been used to select a complete sample of 656 GPS sources using near simultaneous observations at 4.8, 8.6 and 20GHz. The 361 QSOs and 104 galaxies within this sample are found to be at lower redshifts and lower radio power than those listed in the master list of Labiano et al. (2007), and support the idea that the origins of the GPS spectra are intrinsically different for galaxies and QSOs. Contamination of the GPS sample by variable sources was limited by the near simultaneous nature of the observations, resulting in the largest list of genuine HFP GPS sources so far discovered.
Future work to be done on the sample includes: eVLBI observations of a sample of low redshift, low radio power galaxies to probe their milliarcsecond structure; 40 and 95GHz observations of inverted and HFP galaxies; and a program to expand the number of known redshifts.

\section*{Acknowledgments}
I wish to thank the AT20G team for the use of pre-release data in the publication of this paper, as well as the participants of the 4th CSS/GPS conference who gave their time to discuss and comment on my work and to the SOC for allowing me the opportunity to speak.

This research has made use of the NASA/IPAC Extragalactic Database (NED) which is operated by the Jet Pro-pulsion Laboratory, California Institute of Technology, under contract with the National Aeronautics and Space Administration.
{}

\end{document}